\documentclass[aps,twocolumn,superscriptaddress]{revtex4-1} 

\usepackage{graphicx}  
\usepackage{dcolumn}   
\usepackage{bm}        
\usepackage{amssymb}   
\usepackage{subfigure}
\usepackage{epstopdf}
\usepackage{comment}
\usepackage{color}
\usepackage{slashed}
\usepackage[utf8]{inputenc}
\usepackage{multirow}
\usepackage{footnote}
\usepackage{slashed}
\usepackage{amsmath}
\allowdisplaybreaks[4]
\usepackage{accents}
\newlength{\dhatheight}

\usepackage{hyperref}

\def\figureautorefname~#1\null{Fig.\,#1\null}
\def\tableautorefname~#1\null{Tab.\,#1\null}

\def\equationautorefname~#1\null{Eq.\,(#1)\null}

\begin{document}

\title{Search for boosted dark matter with high-Z material in underground experiments}
\author{ Xin Chen }
\email{xin.chen@cern.ch}
\affiliation{Department of Physics, Tsinghua University, Beijing 100084, China}
\affiliation{Center for High Energy Physics, Tsinghua University, Beijing 100084, China}
\author{ Litao Yang }
\email{yanglt@mail.tsinghua.edu.cn}
\affiliation{Key Laboratory of Particle and Radiation Imaging (Ministry of Education) and Department of Engineering Physics, Tsinghua University, Beijing 100084, China}


\begin{abstract}
We propose to search for a boosted dark matter (DM) particle from astrophysical sources using an emulsion detector in deep underground facilities. We further propose using high-$Z$ material such as the lead for a larger DM-nucleus coherent scattering cross section above a threshold. The boosted DM will scatter into an excited DM. While the nuclear recoil energy is not detected, the decay products from the excited DM can be recorded by the proposed detector. Backgrounds such as the high energy solar neutrinos can be controlled by the reconstructed track topology. The proposed detector has the potential to find a boosted DM in the sub-GeV mass region, and in certain parameter space, an observation can be made while there is no sensitivity in other large-volume detectors such as Super-K.
\end{abstract}

\maketitle
\newpage



\section{Introduction}

With compelling gravitational evidences for the existence of dark matter (DM) from astrophysical observations, searching for a DM particle either indirectly through their annihilations, or directly through their interactions with target nuclei, have been an endeavor of many experiments. DM particles are usually assumed to be massive and cold, with a typical velocity of $\mathcal{O}$($10^{-3}$). At this low speed, direct detection is very hard since the energy transferred from the DM to a target nucleus is below MeV level, and consequently demands low energy threshold and low radioactivity background in the deep underground experiments. However, in analogy to the Standand Model (SM), it is not impossible that DM sector has multiple components, one of which may have a relativistic speed, so that its detection at an underground experiment is possible. The multi-component DM is used to explain several issues in cosmology. In its simplest form \cite{BDMFlux_1}, it has a heavy secluded component with no direct couplings to the SM particles (so to evade the direct detection), and a light boosted component which interacts with SM, and assists the thermalization of the heavy one with it. Although subdominant in the universe, the light DM can be constantly produced through the annihilation of the heavy ones in the Galactic center, inside the Sun or the Earth \cite{BDMFlux_2}, where the heavy DM's are trapped and have a higher abundance to annihilate.

Such a boosted DM (BDM) has been searched for in the Super-K experiment \cite{SuperK_BDM} with an energy threshold above 100 MeV. This threshold was used to reduce the spallation background induced by muons, and background from neutrino collisions. In \cite{Kim_BDM}, a novel search for a BDM scattering into an excited DM state leading to a three-ring signature (characterized by a proton or electron recoil, and a $e^+e^-$ pair from an excited DM decay) in Super-K was proposed, for a better background suppression. However, when an incoming DM is not energetic enough to ``knock out" a proton or electron, the DM-nucleus collision is elastic and the signature consists of only two rings instead of three. When the DM energy is even lower, interactions will be turned off. The threshold energy for the interaction to happen depends on the nucleus mass, and is lower when the nucleus is heavier. A heavier nucleus has another advantage in that the coherent DM-nucleus scattering cross section scales with $Z^2$, where $Z$ is the atomic number of the target nucleus. In this paper, we propose to use a high-$Z$ material to search for a BDM through its coherent scattering into an excited DM, whose decay products (a $e^+e^-$ pair) can be detected in an underground DM detector. This kind of signal may escape detection in a large volume detector such as Super-K or Hyper-K.

\section{Boosted dark matter energy spectrum}

For simplicity and without losing generality, we assume the two-component fermion DM model with a contact action
\begin{align}
\label{equ:contact_DM}
\frac{\lambda}{\Lambda^2} \bar{\psi}_A\psi_A\bar{\psi}\psi,
\end{align}
where $\psi_A$ ($\psi$) is the dominant heavy (subdominant light) DM fermion with mass $m_A$ ($m$), $\lambda$ is the effective coefficient and $\Lambda$ the energy scale below which the effective operator of Eq. \ref{equ:contact_DM} is valid. When the velocity of $\psi_A$ is non-relativistic, the $\psi$'s energy is almost a delta function, $\delta(E-m_A)$, jittered by each $\psi_A$'s thermal velocity. Suppose the velocities of the two initial state DM particles are $\boldsymbol{\upsilon_1}$ and $\boldsymbol{\upsilon_2}$, then up to the first power of these velocities, the outgoing $\psi$'s energy can be expressed as
\begin{align}
\label{equ:energy_BDM}
E = m_A + \sqrt{m_A^2-m^2} \upsilon_s \cos\theta,
\end{align}
where $\boldsymbol{\upsilon_s} = (\boldsymbol{\upsilon_1}+\boldsymbol{\upsilon_2})/2$ is the total velocity of the $\psi_A\psi^*_A$ system, $\theta$ is the polar angle between $\boldsymbol{\upsilon_s}$ and the velocity $\boldsymbol{\upsilon}=(\boldsymbol{\upsilon_1}-\boldsymbol{\upsilon_2})/2$ of the outgoing $\psi$ in the center-of-mass (CM) frame of the $\psi_A\psi^*_A$ pair. The differential production rate of $\psi$ is proportional to
\begin{align}
\label{equ:prob_BDM}
\sigma \upsilon f(\boldsymbol{\upsilon_1}) d^3 \upsilon_1 f(\boldsymbol{\upsilon_2}) d^3 \upsilon_2,
\end{align}
where $\sigma$ is the $\psi_A\psi^*_A$ annihilation cross section, $f(\boldsymbol{\upsilon})=\pi^{-3/2}\upsilon_0^{-3}e^{-\upsilon^2/\upsilon_0^2}$ is the probability function defined in the DM Standard Halo Model \cite{SHM}, with $\upsilon_0\simeq 235$ km/s being the most probable speed \cite{SHM_2}. The $\sigma \upsilon$ in Eq. \ref{equ:prob_BDM} is nearly a constant, so the energy spread of $\psi$ around $E=m_A$ can be calculated as
\begin{align}
\label{equ:dE_BDM}
\overline{(\Delta E)^2} & = \frac{m_A^2-m^2}{2} \int \upsilon_s^2 \cos^2\theta f(\boldsymbol{\upsilon_1}) f(\boldsymbol{\upsilon_2}) d^3 \upsilon_1 d^3 \upsilon_2 d\cos\theta, \nonumber \\
 \Delta E & = \sqrt{\left(\frac{1}{4}+\frac{2}{3\pi} \right) \left(m_A^2 - m^2\right)} \upsilon_0.
\end{align}
To give a typical estimation, for $m_A=50$ MeV and $m=10$ MeV, $\Delta E$ is about 26 keV.

\section{Mass splitting for the light DM}

We assume the following realization of light dark fermion $\psi$ mass splitting
\begin{align}
\label{equ:lagr}
\mathcal{L} \supset & - \frac{1}{4} X_{\mu\nu}X^{\mu\nu} + \frac{1}{2} m^2_X X_\mu X^\mu - \epsilon e Q_f X_\mu \bar{f}\gamma^\mu f +\nonumber \\
& \bar{\psi} i\gamma^\mu D_\mu \psi - m_D \bar{\psi}\psi - \frac{1}{2} m_M\left( \bar{\psi^c}\psi + \bar{\psi} \psi^c \right),
\end{align}
where $X$ is a dark photon mediating the force between the usual and dark sectors, $f$ is the SM fermion with charge $Q_f$, $\epsilon$ is the mixing parameter between the $U(1)_Y$ and $U(1)_D$ gauge fields, $D_\mu=\partial_\mu + ig_D X_\mu$ is the covariant derivative with dark coupling parameter $g_D$, $\psi$ is the light fermion in Eq. \ref{equ:contact_DM}, and $\psi^c$ is its charge-conjugate field. The dark photon mass term may result from a spontaneous $U(1)_D$ breaking, whose detailed realization is not specified here. The last two terms in Eq. \ref{equ:lagr} are the Dirac and Majorana mass terms, with the latter explicitly violating $U(1)_D$ and causing a mass splitting for $\psi$. To see this, substituting $(\psi_{1}\pm\psi_{2})/\sqrt{2}$ (where $\psi_{1,2}$ are two Majorana fermion fields) for $\psi$ and $\psi^c$ respectively to obtain
\begin{align}
\label{equ:lagr2}
 \mathcal{L} \supset & \frac{1}{2} \bar{\psi}_1 \left( i \slashed{\partial} - m_1 \right) \psi_1  + \frac{1}{2} \bar{\psi}_2 \left( i \slashed{\partial} - m_2 \right) \psi_2 \nonumber \\
 & - g_D \bar{\psi}_1 \gamma^\mu \psi_2 X_\mu,
\end{align}
where $m_1 = m_D+m_M$ and $m_2 = m_D-m_M$. It is evident that two Majorana dark fermions emerge with different masses. It is also possible to achieve a similar mass splitting for a scalar dark matter by the presence of a similar $U(1)_D$ violating mass term \cite{Excited_DM}, but the fermion model we are studying will be general enough to cover similar kinematics. 

We consider the mass degeneracy case where $m_A\simeq m_1$, so that $\psi_A$ predominantly decays into $\psi_2$, and the cases in which the excited state $\psi_1$ can be hardly produced at Super-K.

\section{Boosted dark matter scattering into excited dark matter}

When the energy transfer between $\psi_2$ and nucleus is about a few tens of MeV, the scattering is elastic and we assume the target atom is a scalar boson (its spin is not important). The vertex read
\begin{align}
\label{equ:vertex}
ieF(q^2)(P_i+P_f)_\mu,
\end{align}
where $q^2$ is the momentum transfer squared, $P_{i,f}$ are the target atom's initial and final state 4-vectors, $F(q^2)$ accounts for the $q^2$-dependent form factors. 

\begin{figure}[!htb]
\centering
\includegraphics[width=0.35\textwidth]{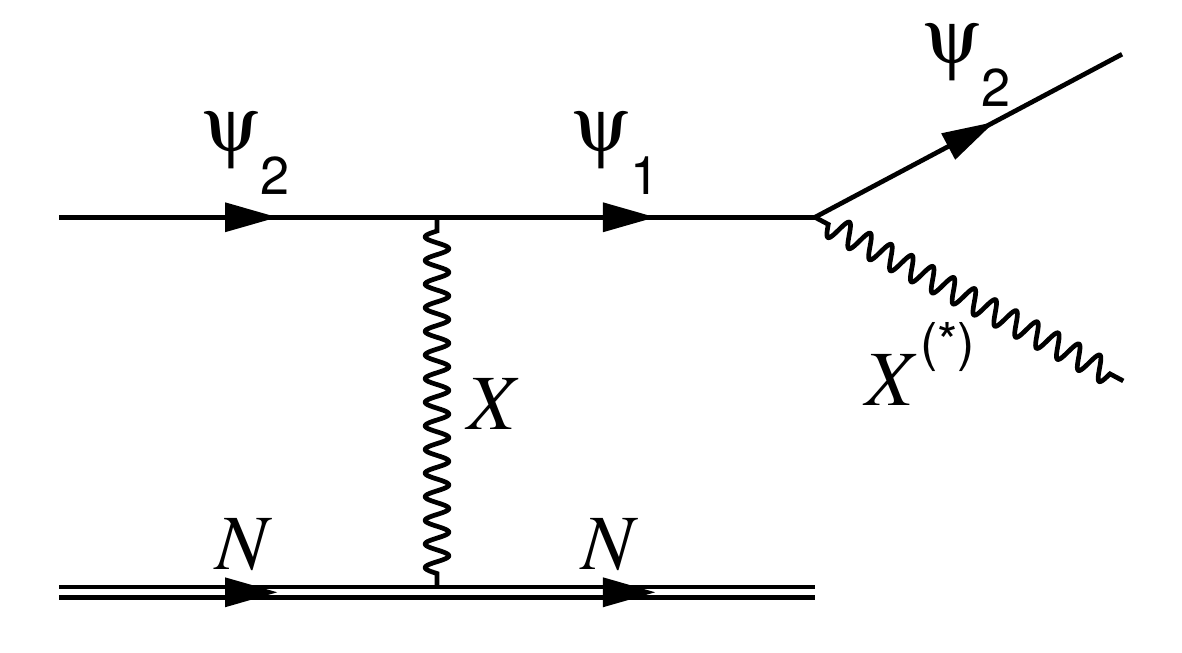}
\caption{The process of boosted dark matter particle $\psi_2$ scattering off a nucleus $N$ into an excited dark matter particle $\psi_1$. The latter then decays back into $\psi_2$ and $X$ (either on or off-shell). The decay product of $X$ ($e^+e^-$) will be then detected. }
\label{fig:feyn}
\end{figure}

As illustrated in Fig. \ref{fig:feyn}, the scattering process is $\psi_2(E_2)+N(m_N) \to \psi_1(E_1)+N(E_N)$, where $N$ denotes the target atom with mass $m_N$ and the particle's energy is indicated in the parentheses. The differential scattering cross section can be then expressed as
\begin{align}
\label{equ:xs}
\frac{d\sigma_\text{scat.}}{dt} = & \frac{\alpha \epsilon^2 g_D^2 G_2(t)}{4 m_N^2\left(E_2^2-m_2^2\right)\left(m_X^2+t\right)^2} \left[ \left( 4m_N E_2  -m_1^2 + \right. \right.  \nonumber \\
& \left. \left. m_2^2 -t\right)^2 - \left( 4m_N^2 +t \right) \left( \left(m_1-m_2\right)^2 + t \right) \right],
\end{align}
where $\alpha$ is the fine structure constant, $t=-q^2=2m_N(E_2 - E_1)\geq 0$, and $G_2(t)$ is an overall elastic form factor defined as in \cite{FF_1,FF_2,FF_3},
\begin{align}
\label{equ:G2}
G_2(t) = \left(\frac{a^2 t}{1+a^2 t}\right)^2 \left(\frac{1}{1+t/d}\right)^2 Z^2,
\end{align}
where the first term accounts for the atomic form factor due to electron screening, and the second one for the nuclear elastic form factor. According to the simple parameterization in \cite{FF_1}, $a=111.7 Z^{-1/3}/m_e$, and $d=0.164 A^{-2/3}$ GeV$^2$ with $A$ being the target atomic mass number. There is also an inelastic scattering part whose contribution is small and can be neglected at low momentum transfer, since we are focusing on the enhanced coherent scattering. The upper and lower bounds for $t$ read
\begin{align}
\label{equ:bound_of_t}
t^\pm = & \frac{m_N}{s} \left[ \left(2E_2^2 - m_1^2 - m_2^2 \right)m_N - \left(m_1^2-m_2^2\right)E_2 \right. \nonumber \\
& \left. \pm\sqrt{(E_2^2-m_2^2)\lambda}\right], 
\end{align}
where $s = 2m_N E_2 + m_N^2 + m_2^2$ is the CM energy of the initial $\psi_2+N$ system, and $\lambda = \left(2m_N E_2 - m_1^2 + m_2^2 \right)^2 - 4m_1^2 m_N^2$. The cross section decreases rapidly as $t$ increases, as Eq. \ref{equ:xs} indicates, but it is the range of low $t$ values that gives the most important contributions. 
The total scattering cross section can be obtained by integrating Eq. \ref{equ:xs} over the range of $t$ determined by Eq. \ref{equ:bound_of_t}.

\begin{figure}[!htb]
\centering
\includegraphics[width=0.40\textwidth]{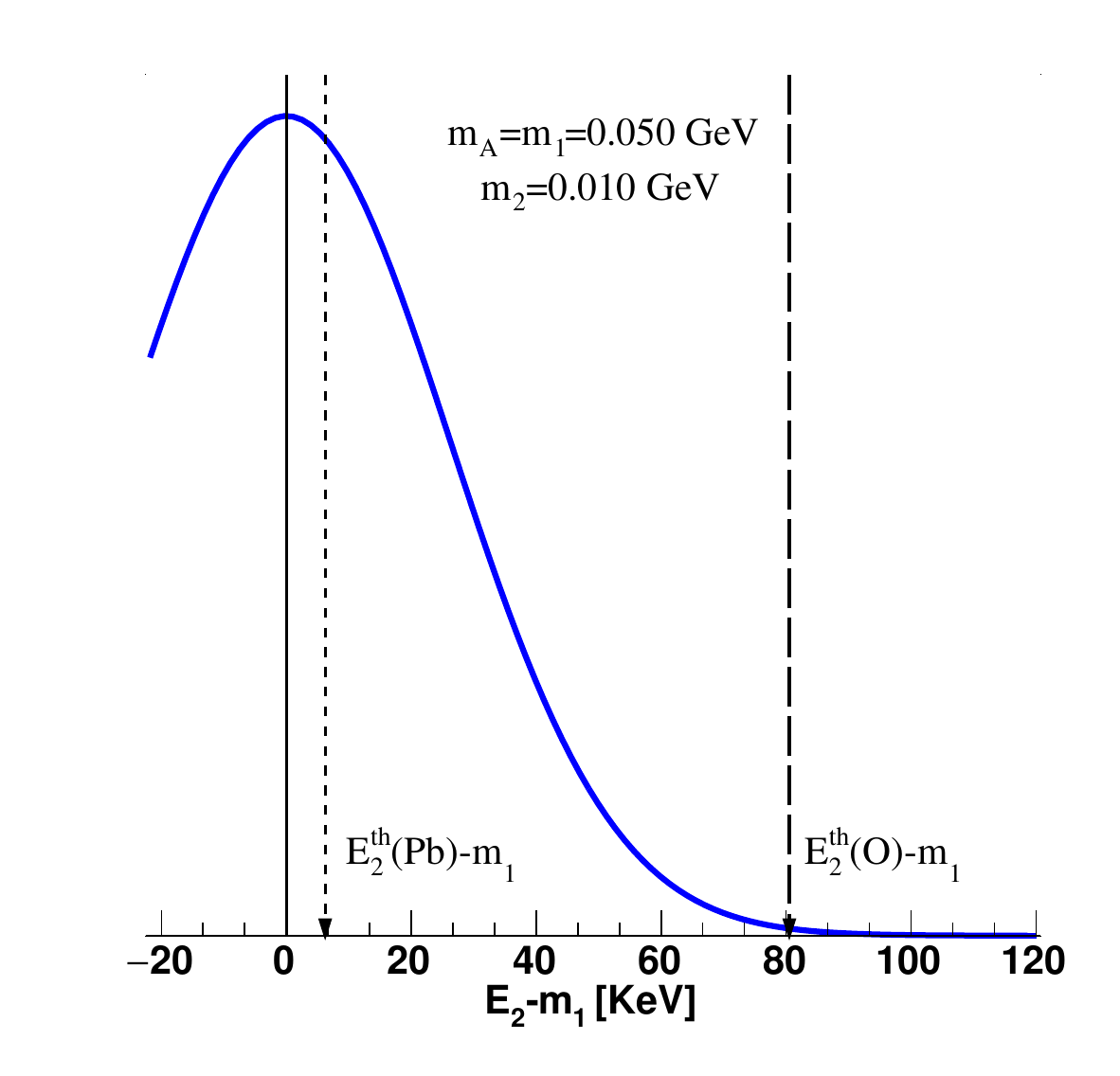}
\caption{The probability distribution of incoming $\psi_2$'s energy (blue Gaussian curve) and the scattering energy thresholds for Pb and Oxygen atoms represented by the dashed arrows, for a particular set of parameter values indicated in the plot.}
\label{fig:threshold}
\end{figure}

For the scattering process to happen, the minimum threshold energy required for the incoming $\psi_2$ is
\begin{align}
\label{equ:E2min}
E_2^\text{th} = m_1 + \frac{m_1^2-m_2^2}{2m_N},
\end{align}
from which it is evident that, different from the usual elastic scattering process where lighter nucleus mass is favored for larger nuclear recoil energy (as in the contact interaction models), in the case of excited DM, there is a threshold energy that decreases with larger nucleus mass. As a result, high-$Z$ materials are more effective to detect this type of DM scattering than low-$Z$ ones. In Fig. \ref{fig:threshold}, the probability distribution of incoming $\psi_2$'s energy according to Eq. \ref{equ:dE_BDM}, and the scattering energy thresholds for lead (Pb) and Oxygen (O) target atoms according to Eq. \ref{equ:E2min}, are shown for $m_A=m_1=50$ MeV and $m_2=10$ MeV. The Oxygen is the main target atom of the purified water at Super-K (the contribution from hydrogen atoms can be neglected). It is evident that in the case of $m_A\lesssim m_1$, due to the higher threshold, the available BDM flux above it is severely suppressed.

\section{Decay of the excited dark matter}

Depending on the $\psi_1$'s mass $m_1$, the decay process $\psi_1 \to\psi_2 e^+ e^-$ proceeds via either an on-shell (if $m_1>m_2+m_X$) or an off-shell (if $m_1<m_2+m_X$) dark photon $X$. For the off-shell $X^*$ three-body decay of $\psi_1$, the partial decay width reads
\begin{align}
\label{equ:psi1_wid_2}
\frac{d\Gamma_{\psi_1}}{dm_{ee}} = & \frac{\alpha \epsilon^2 g_D^2 m_{ee} (2m_e^2+m_{ee}^2) }{6\pi^2 m_1^3 (m_X^2-m_{ee}^2)^2} \left[ \left(m_1^2-m_2^2\right)^2 - 2 m_{ee}^4+ \right.  \nonumber \\
& \left. \left(m_1^2+m_2^2-6m_1 m_2\right)m_{ee}^2 \right] \left(1-\frac{4m_e^2}{m_{ee}^2}\right)^{\frac{1}{2}} \cdot \nonumber \\
& \left[ \left(\frac{m_1^2 - m_2^2}{m_{ee}^2}-1\right)^2 - \frac{4m_2^2}{m_{ee}^2} \right]^{\frac{1}{2}},
\end{align}
where $m_{ee}$ denotes the $e^+e^-$ pair's invariant mass, whose allowed range is $2m_e\leq m_{ee}\leq (m_1-m_2)$. 


When $m_1>m_2+m_X$, $\psi_1$ decays promptly due to the large values of $g_D$. The total decay width of $\psi_1$ reads
\begin{align}
\label{equ:psi1_wid}
\Gamma(\psi_1\to\psi_2 X)  = & \frac{g_D^2}{4\pi m_1} \left[ m_1^2+m_2^2-6m_1 m_2-2m_X^2 + \right. \nonumber \\
& \left. \frac{\left(m_1^2-m_2^2\right)^2}{m_X^2}\right] \left[ 1-\frac{(m_2+m_X)^2}{m_1^2}\right]^{\frac{1}{2}} \cdot \nonumber \\
&  \left[ 1-\frac{(m_2-m_X)^2}{m_1^2}\right]^{\frac{1}{2}} .
\end{align}
And the total decay width of $X\to e^+e^-$ can be expressed as
\begin{align}
\label{equ:DPh_wid}
\Gamma(X\to ee)  = & \frac{1}{3} \alpha \epsilon^2 m_X \left(1+\frac{2m_e^2}{m_X^2} \right)  \left(1-\frac{4m_e^2}{m_X^2} \right)^{\frac{1}{2}}.
\end{align}
The decay lifetime times the speed of light $c\tau$ for $\psi_{1,2}$ for several different sets of parameters is listed in Tab. \ref{tab:decay_ctau}.  The largest $c\tau$ is found for the off-shell $X$ decay. Considering the fact that $\psi_1$ is created through a collision with a low speed ($\beta\ll 1$), its decay length is well below a mm level. Therefore, $\psi_{1,2}$ effectively decay promptly at the moment they are created.

\begin{table}[!hbt]
\centering
\caption{ The decay lifetime times the speed of light $c\tau$ for several different sets of parameters together with $\epsilon=10^{-3}$ and $g_D=1$.}
\label{tab:decay_ctau}
\begin{tabular}{c|cc}
\hline\hline
 ($m_A=m_1,m_2,m_X$),  & $c\tau$($\psi_1$) & $c\tau$($\psi_2$) \\ 
 all in MeV & mm & mm \\ \hline
 ($50,10,50$) & 0.14 & --- \\ \hline
 ($50,10,20$) & $1.2\times 10^{-11}$ & $4.1\times 10^{-3}$ \\ \hline
 ($75,10,20$) &  $2.7\times 10^{-12}$ & $4.1\times 10^{-3}$ \\ \hline
 ($75,10,50$) &  $4.3\times 10^{-11}$ & $1.6\times 10^{-3}$ \\ \hline
 ($100,10,20$) & $1.1\times 10^{-12}$ & $4.1\times 10^{-3}$ \\ \hline
 ($100,50,30$) & $9.8\times 10^{-12}$ & $2.7\times 10^{-3}$ \\ \hline
 ($150,100,30$) & $7.4\times 10^{-12}$ & $2.7\times 10^{-3}$ \\ \hline
\hline
\end{tabular}
\end{table}

\section{Detecting boosted DM with excited DM decay}

\begin{figure}[!htb]
\centering
\includegraphics[width=0.35\textwidth]{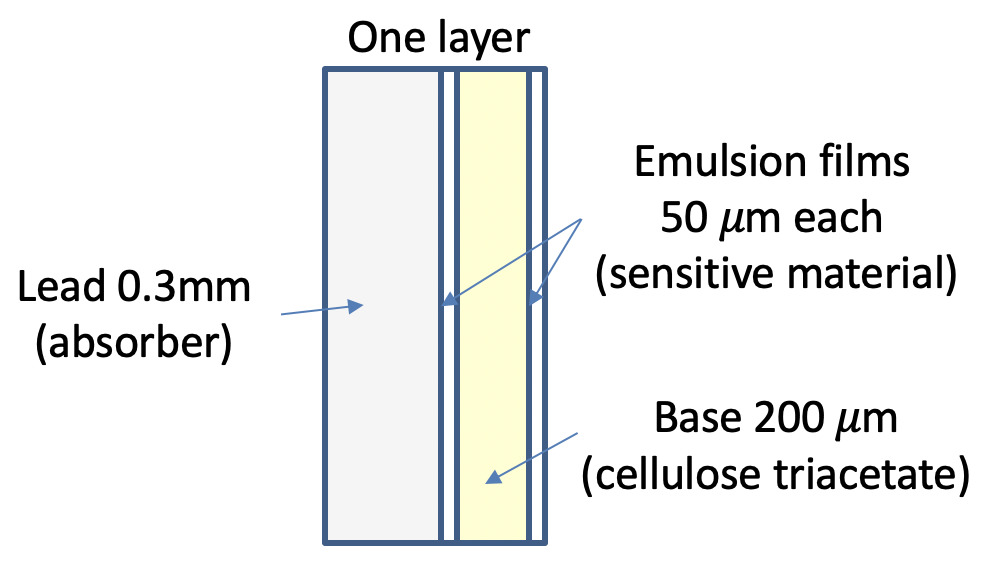}
\caption{The illustration of one emulsion layer of the proposed detector.}
\label{fig:emulsion_layer}
\end{figure}

To detect the $e^+e^-$ pair from the excited DM decay after the scattering of a cosmological BDM with a target nucleus, we propose to use an emulsion detector with Pb as the target and absorber material at the same time. Emulsion detectors have been used in the Opera \cite{Opera}, DsTau \cite{DsTau} and FASER$\nu$ \cite{FASERnu} experiments. We propose the dimension of a subdetector to be 30$\times$30$\times$108 cm$^3$, consisting of 1800 layers, with each layer 0.6 mm think. Each layer is comprised of a 0.3 mm thick lead plate and 0.3 mm thick emulsion layer. Each emulsion layer consists of a 200 $\mu$m thick base (made of, e.g., cellulose acetate), sandwiched between two emulsion films of 50 $\mu$m thickness, as illustrated in Fig. \ref{fig:emulsion_layer}. The emulsion films consist mainly of AgBr (about 66\%) and gelatin material (34\%). The silver bromide crystals are sensitive to ionization by charged particles passing through it (with an energy band gap of 2.5 eV), and have a typical size of 0.2 $\mu$m. Therefore, the emulsion can have a position measurement of tracks with a precision below 1 $\mu$m, which makes it ideal for our purpose of detecting a pair of $e^+e^-$ with an energy above $\sim$5 MeV with good tracking performance. We propose to make 9 identical subdetectors, with each one having a cross section area of 30$\times$30 cm$^2$. The total sensitive target mass will be about 5 tons with 0.875 m$^3$ effective volume. After particle events are recorded by the emulsion detector for an accumulated period, the films will be developed and the AgBr grains positions will be read out by dedicated microscopes. The film on each side of the emulsion layer can provide a position measurement. At the reconstruction level, sequences of aligned grains will be recognized and form tracks for the electrons.

The lead is used instead of, e.g. tungsten, apart from its high-$Z$, is because it is relatively inexpensive, and an electron can travel longer to form a longer track.
The main radioactive contamination in lead is $^{210}$Pb, which decays into $^{210}$Bi and $^{210}$Po. There is a 1.16-MeV $\beta$ ray from $^{210}$Bi, but electron of this energy can hardly travel more than one detector layer according to our simulation, and the $^{210}$Pb contamination can be controlled in the production pathway.

We propose to place the detector in a deep underground facility, so the cosmic muon background can be suppressed as much as possible. On the other hand, high energy muons going through the detector will cause a long track from one side of the detector to another side, and can be more easily identified and removed than other experiments that have not as much precise tracking ability, such as Super-K. 

\begin{figure}[!htb]
\centering
\includegraphics[width=0.40\textwidth]{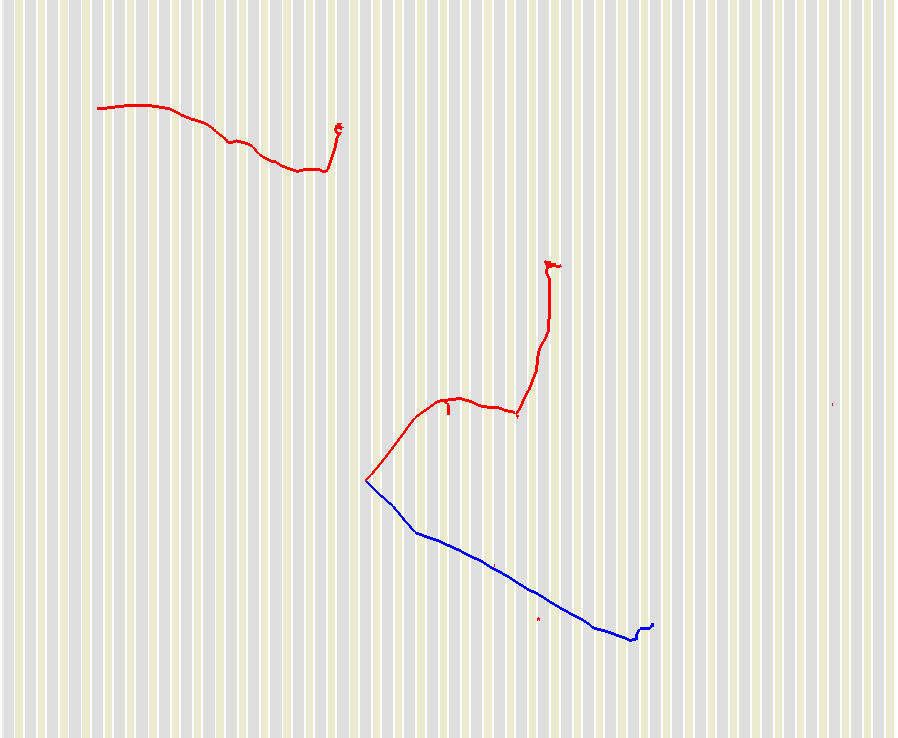}
\caption{The 2-D view of the 3-D track of a 10 MeV single electron (upper left), and those for a $e^+e^-$ pair with an energy of 15 MeV each (bottom right), in part of the emulsion detector as simulated by Geant4 \cite{Geant4}. The electron (positron) trajectory is shown in red (blue). The lead plate (in light grey) and the emulsion base (in light yellow)  in each layer are visible. }
\label{fig:geant4}
\end{figure}

Solar neutrino is an important background for this search. Most solar neutrinos have energy below 2 MeV, except for those from the $^8$B reaction chain that extends to about 13 MeV. The most serious interaction for our search is $\nu_e n\to e^- p$, with a cross section of about $10^{-41}$ cm$^2$. The total solar neutrino flux with $E_\nu \gtrsim10$ GeV is about $10^5$ cm$^{-2}$s$^{-1}$ \cite{SolarNeutrino}, taking into the neutrino oscillation effect. The proposed detector will see about 60 such high energy neutrino interactions per year. Figure \ref{fig:geant4} shows a 2-D view of the 3-D track of a 10 MeV single electron (from neutrino background) and those for a $e^+e^-$ pair with an energy of 15 MeV each\footnote{There are often bremsstrahlung photons from the electron or positron that deposit little energy in the emulsion films. Thus, they are not shown in Fig. \ref{fig:geant4}.}, simulated by Geant4 \cite{Geant4}. The two cases can be differentiated by the the total length of the trajectory, and in the case of the $e^+e^-$ pair, a ``kink" in the middle of the trajectory. An offline analysis exploiting MVA techniques is expected to  separate better the signal and background.

Finally, the solar neutrino interaction cross section for $\nu_e e^- \to \nu_e e^-$ is two orders of magnitude lower than for $\nu_e n\to e^- p$, which has a large energy spread for the outgoing electron. The atmospheric neutrinos are many orders of magnitude lower than the solar ones. Therefore, these two backgrounds are neglected in this search.

\section{Signal sensitivity}

The total signal yield can be calculated as
\begin{align}
\label{equ:sig_yield}
N_{\text{sig}} = \mathcal{L} t \sigma_\text{scat.} N_T,
\end{align}
where $\mathcal{L}$ is the BDM flux, $t$ is the exposure time and $N_T$ is the number of target nuclei. An integration over a probability function is needed when $\mathcal{L}$ is energy dependent like in Fig. \ref{fig:threshold}. Some predictions for $\mathcal{L}$ is at $\mathcal{O} (10^{-7}-10^{-4})$ cm$^{-2}$s$^{-1}$ \cite{BDMFlux_1,BDMFlux_2,BDMFlux_3}. When we take $\mathcal{L}=10^{-4}$ cm$^{-2}$s$^{-1}$ which is Gaussian distributed with a resolution according to Eq. \ref{equ:dE_BDM}, $\epsilon=10^{-3}$ and $g_D=1$, the predicted signal yields with 5 years' exposure time for the proposed detector and Super-K, are given in Tab. \ref{tab:sens}. It can be seen that smaller $m_X$ leads to larger yields, and larger mass splitting ($m_1-m_2$) leads to smaller yields. However, the signal with larger mass splitting will also give more outstanding signatures in the emulsion detector (more energetic $e^+e^-$ pair). Moreover, as either or both of $m_1$ and $m_2$ increase, the signal yields also decrease, and the expected yields at Super-K diminish towards zero while an observation is still possible with the proposed detector.

\begin{table}[!hbt]
\centering
\caption{ The expected BDM signal yields with 5 years' exposure time for the proposed emulsion detector and Super-K, for several different sets of parameters together with $\epsilon=10^{-3}$, $g_D=1$, $\mathcal{L}=10^{-4}$ cm$^{-2}$s$^{-1}$ and Gaussian distributed with an energy resolution $\Delta E$ according to Eq. \ref{equ:dE_BDM}}. 
\label{tab:sens}
\begin{tabular}{c|cc}
\hline\hline
 ($m_A=m_1,m_2,m_X$),  & \multicolumn{2}{c}{$N_{\text{sig}}$} \\ \cline{2-3}
 all in MeV & Emul. Det. & Super-K \\ \hline
 ($50,10,50$) & 19.8 & 28.6 \\ \hline
 ($50,10,20$) & 60.8 & 87.8 \\ \hline
 ($75,10,20$) &  11.9 & 0.04 \\ \hline
 ($75,10,50$) & 6.5 & 0.02 \\ \hline
 ($100,10,20$) & 2.8 & $2.0\times 10^{-6}$ \\ \hline
 ($100,50,30$) & 9.7 & $7.1\times 10^{-4}$ \\ \hline
 ($150,100,30$) & 4.4 & $2.0\times 10^{-8}$ \\ \hline
\hline
\end{tabular}
\end{table}

\begin{figure}[!htb]
\centering
\includegraphics[width=0.40\textwidth]{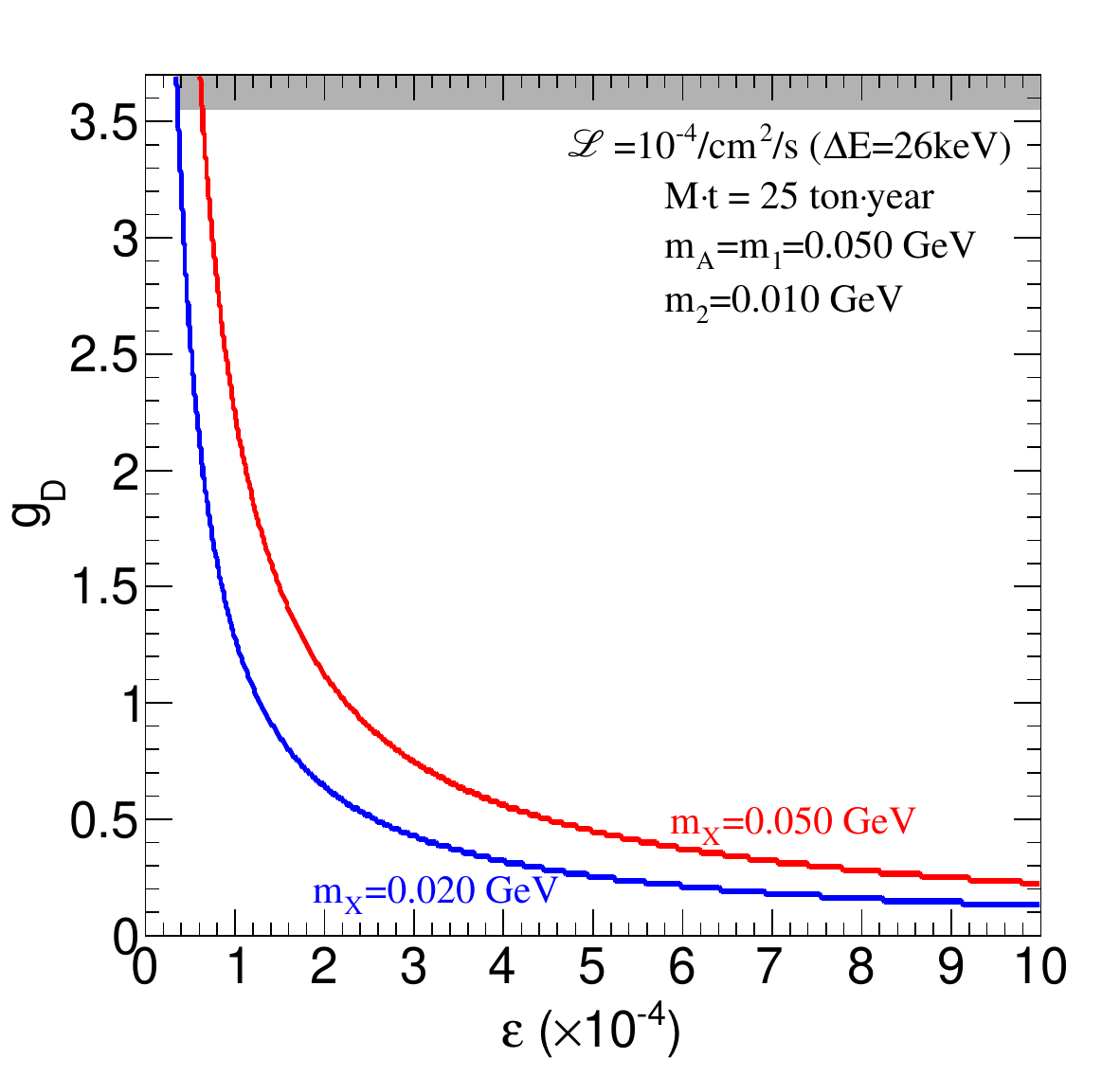}
\caption{The parameter space in $\epsilon$ vs. $g_D$ with at lease one expected signal event for $\mathcal{L}=10^{-4}$ cm$^{-2}$s$^{-1}$, $m_A=m_1=50$ MeV, $m_2=10$ MeV (with $\Delta E=26$ keV), and two different values for $m_X$. The grey region is where the perturbativity fails due to too large $g_D$. }
\label{fig:sens1}
\end{figure}

\begin{figure}[!htb]
\centering
\includegraphics[width=0.40\textwidth]{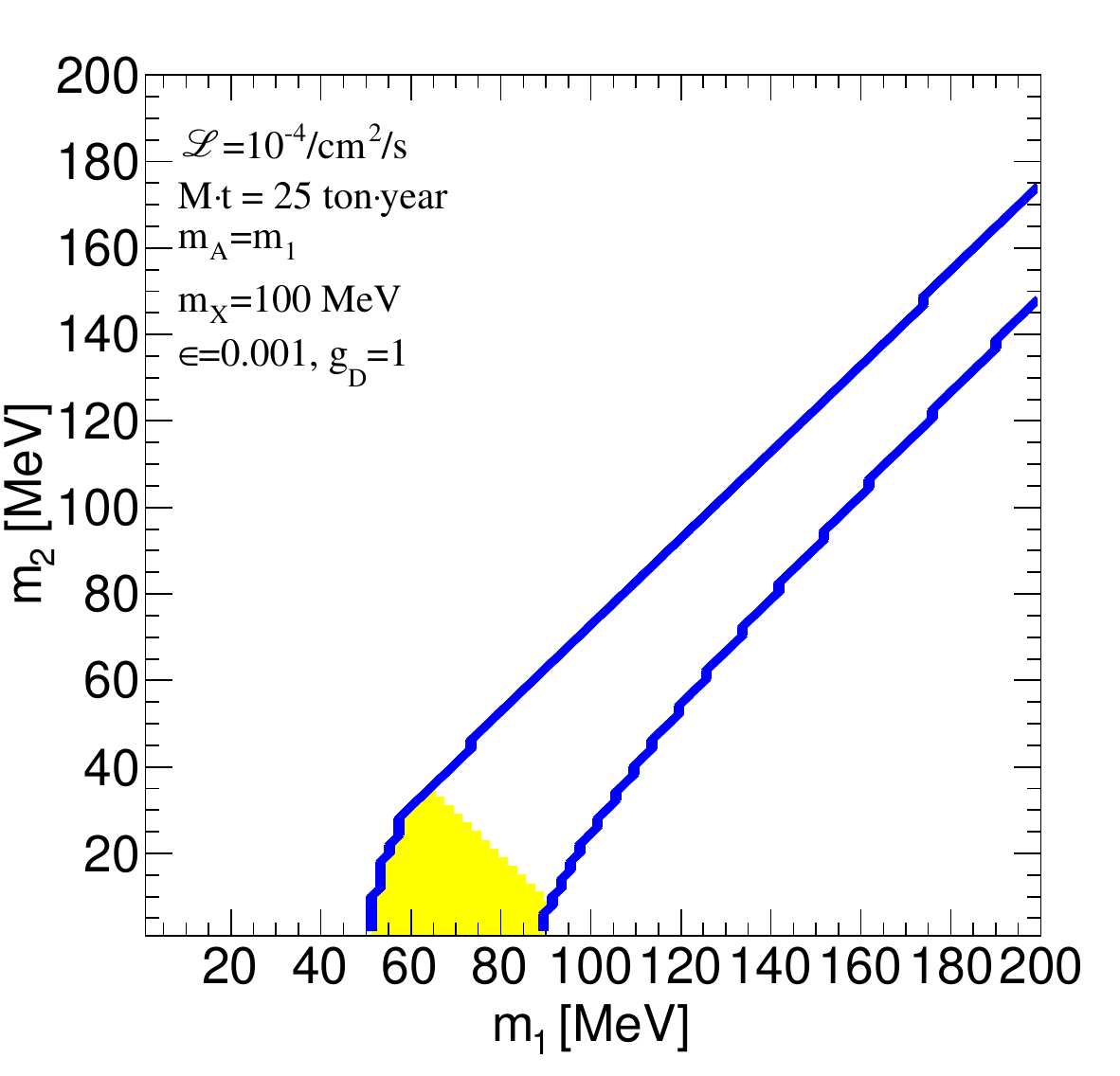}
\caption{The parameter space in $m_1$ vs. $m_2$ with at lease one expected signal event for $\mathcal{L}=10^{-4}$ cm$^{-2}$s$^{-1}$, $m_A=m_1$, $\epsilon=10^{-3}$, $g_D=1.0$, $m_X=100$ MeV, $c\tau<1$ cm for $\psi_1$, and the requirement that the signal yield overtakes that at Super-K. The yellow region is where the decay chain of $X\to\psi_1\psi_2, \psi_1\to\phi_2 e^+ e^-$ is open. }
\label{fig:sens2}
\end{figure}

Finally, the parameter space with at lease one signal event expected is shown in Fig. \ref{fig:sens1} for $\mathcal{L}=10^{-4}$ cm$^{-2}$s$^{-1}$, $m_A=m_1=50$ MeV, $m_2=10$ MeV (with $\Delta E=26$ keV), and two different values for $m_X$. It is seen that there is a large parameter space to be explored. The grey area is where $g_D>\sqrt{4\pi}$ and perturbativity fails. On the other hand, in Fig. \ref{fig:sens2}, the values for $\epsilon$, $g_D$ and $m_X$ are fixed, and those for $m_{1,2}$ are varied. The region with at least one signal event, $c\tau<1$ cm for $\psi_1$, and the requirement that the signal yield with the emulsion detector is more than that expected at Super-K, is enclosed by the blue contour. The yellow region is interesting, because it is where the decay chain of $X\to\psi_1\psi_2, \psi_1\to\phi_2 e^+ e^-$ is open. In the direct searches for a relatively heavy dark photon at a $e^+e^-$ collider \cite{BaBar_Atoll}, usually $X$ decaying into dark matter is assumed forbidden. The yellow region in Fig. \ref{fig:sens2} can avoid the dark photon limits set by such experiments. 

\section{Conclusion}

In this paper, we propose an emulsion detector placed in deep underground facilities to search for a boosted dark matter from various astrophysical sources (from dark matter annihilation in the Galactic center, Sun, Earth, or even the cosmic rays \cite{cosmic_upscatter}). We further propose a high-$Z$ material such as Pb as the target, which coherently scatters the BDM into an excited DM, whose decay product (a $e^+e^-$ pair) can be recorded by the detector (in this case the nuclear recoil is too small to be detected). We assumed a Gaussian distributed BDM energy spectrum, and then estimated the signal yields with different parameter settings. We found that the proposed detector can detect a BDM with a sub-GeV mass, and in some cases the signal has no chance to be found in Super-K at all while an observation in the proposed detector is possible. Some parameter space explored in this model can escape the limits from direct searches at a $e^+e^-$ collider. The largest background for this search is the high energy solar neutrinos, but can be controlled by track topologies reconstructed in the emulsion detector. The proposed search will certainly complement the existing DM search experiments and deepen our understanding of the dark matter physics.


\begin{acknowledgments}
X. Chen is supported by Tsinghua University Initiative Scientific Research Program. 
\end{acknowledgments}

\bibliographystyle{bibsty}
\bibliography{references}

\end{document}